# Hardware Realization of Neuromorphic Computing with a 4-Port Photonic Reservoir for Modulation Format Identification

Enes Şeker[1,2], Rijil Thomas[1], Guillermo von Hünefeld[3,4], Stephan Suckow[1,*], Mahdi Kaveh[3], Gregor Ronniger[3], Pooyan Safari[3], Isaac Sackey[3], David Stahl[5], Colja Schubert[3], Johannes Karl Fischer[3], Ronald Freund[3,4], and Max C. Lemme[1,2,*]

(1) AMO GmbH, Advanced Microelectronic Center Aachen (AMICA), Otto-Blumenthal-Straβe 25, 52074 Aachen, Germany

(2) Chair of Electronic Devices, RWTH Aachen University, Otto-Blumenthal-Straße 25, 52074 Aachen, Germany

(3) Fraunhofer Institut für Nachrichtentechnik, Heinrich Hertz Institute (HHI), Einsteinufer 37, 10587 Berlin, Germany

(4) Technical University of Berlin, Photonic Communication Systems, Straße des 17. Juni 135, 10623 Berlin, Germany

(5) ID Photonics GmbH, Anton-Bruckner-Straße 6, 85579 Neubiberg, Germany

*Email: suckow@amo.de; max.lemme@rwth-aachen.de

**Abstract**

The fields of machine learning and artificial intelligence drive researchers to explore energy-efficient, brain-inspired new hardware. Reservoir computing encompasses recurrent neural networks for sequential data processing, matches the performance of other recurrent networks with less training and lower costs. However, traditional software-based neural networks suffer from high energy consumption due to computational demands and massive data transfer needs. Photonic reservoir computing overcomes this challenge with energy-efficient neuromorphic photonic integrated circuits or NeuroPICs. Here, we introduce a reservoir NeuroPIC used for modulation format identification in C-band telecommunication network monitoring. It is built on a silicon-on-insulator platform with a 4-port reservoir architecture consisting of a set of physical nodes connected via delay lines. We comprehensively describe the NeuroPIC design and fabrication, experimentally demonstrate its performance, and compare it with simulations. The NeuroPIC incorporates non-linearity through a simple digital readout and achieves close to 100% accuracy in identifying several configurations of quadrature amplitude modulation formats transmitted over 20 km of optical fiber at 32 GBaud symbol rate. The NeuroPIC performance is robust against fabrication imperfections like waveguide propagation loss, phase randomization, etc. and delay line length variations. Furthermore, the experimental results exceeded numerical simulations, which we attribute to enhanced signal interference in the experimental NeuroPIC output. Our energy-efficient photonic approach has the potential for high-speed temporal data processing in a variety of applications.

1. Introduction

Neuromorphic Engineering is an interdisciplinary attempt to match computing hardware to the algorithmic elements of machine learning (ML) and artificial intelligence (AI), thereby reaching faster and more energy-efficient information processing capabilities [1–4]. Recurrent neural networks (RNN) are a subtype of artificial neural networks (ANN), specifically designed to process sequential data such as time series or text data. Unlike feedforward neural networks, which process input data in a single pass, RNNs can maintain a hidden state that allows them to process sequences of inputs, one element at a time. However, since RNNs feature complicated feedback loops that are hard to train, reservoir computing (RC) emerged as a successful alternative to RNNs that simplifies the training process by allowing signals to mix inside a fixed passive reservoir circuit and only requires the output to be trained [1,3,5–8] (see schematic in Figure 1a. The reservoirs cannot and need not be trained to perform the transformation of the input data to the readout layer. The readout part of the system contains non-linear elements, where the output can be extracted in high-dimensional space from the linear input and trained accordingly. By leveraging the strengths of both photonics and RNNs, photonic reservoir computing (PRC) offers a promising hardware-accelerated solution for processing temporal data with high speed and low power consumption. The use of ANNs to solve telecommunication network problems has increased in recent years, primarily due to their energy-efficient operation. PRC combines the benefits of RNNs with photonics, such as large bandwidths, the option for massive parallelism via wavelength multiplexing, and further practical advantages, such as immunity to electromagnetic interference and the possibility of co-integration with microelectronics [3,5,9].

One of the earliest implementations of PRC was proposed by Vandoorne et al. [10,11] who showed through simulation that a network of coupled semiconductor optical amplifiers (SOAs) can be used as a reservoir to recognize slow speech signals. It was shown that a PRC chip can perform well on several benchmark problems despite certain challenges compared to a software counterpart, like an inability to implement negative weights [1]. Other simulation studies have also been conducted in this field, such as PRC of coupled SOAs to classify noisy time series [11], micro-ring resonator (MRR) node-based reservoirs to perform boolean operations like XOR [12] and simple classifications of high bit-rate digital patterns [13], linear passive multimode interference (MMI) splitter-based reservoirs with specific architectures (named swirl [14] and 4-port [3,5,15]) for XOR operations and equalization of nonlinearly distorted signals. While both the swirl and 4-port architecture consist of nodes with inputs and outputs connected to other nodes via delay lines, the 4-Port architecture features an extended network by adding additional connections through the outermost nodes. The MRR-based PRCs produced promising results in classifying digital words with 40 and 160 Gbps bit rates with very low error rates of 0.1% and 0.5%, respectively. This approach enhances the computation efficiency by using the non-linear responses of the MRR reservoir nodes [13]. The versatility of MRRs as the computational unit was also numerically studied in a time-delay-based PRC approach [16] for extracting the time delay signature of optical feedback-induced chaos [17] and for the biomedical application of predicting short to medium ranges of respiratory motions [18].

The 4-port architecture was introduced as an enhancement to the swirl architecture to counter signal loss and improve power distribution among nodes. A simulation study shows that this architecture outperforms the conventional linear feed-forward equalizer in bit error rate (BER) in a 64 quadrature amplitude modulation (QAM) signal for fiber lengths up to 100 km, effectively reducing fiber nonlinearities and transmitter imperfections [5]. Another numerical study utilized a 32 node 4-port architecture to mitigate both linear and nonlinear distortions in optical fiber communication, achieving a BER of 9.14 x 10$^{-5}$ for a 25 Gbps on-off keying (OOK) input signal over a 25 km single-mode transmission [19]. Some numerical studies involve photonic crystals as building blocks to perform boolean tasks, with simulations showing excellent results like XOR operations with bitrates of up to 67 Gbps and header recognition with 6 bits of memory and a bit rate of up to 100 Gbps. However, it must be noted that the fabrication of these blocks remains a challenge with current techniques [20]. In addition to those simulations, several physical implementations of PRC were also shown, such as a 16-node 4x4 swirl architecture on a silicon-on-insulator (SOI) platform [21] and a 32 node version of the 4-port architecture on a SiGe BiCMOS platform [22] both of which are MMI-based and purely passive. The 4x4 swirl architecture performed boolean XOR and NAND operations and a 5-bit header recognition up to 12.5 Gbps. The 32-node architecture on SiGe BiCMOS mitigated non-linear distortions in a 32 Gbps OOK modulated photonic signal and achieved a 0.2 x 10$^{-3}$ forward error correction (FEC) limit.

We have previously modeled and numerically tested the neuromorphic photonic integrated circuits (NeuroPIC) for modulation format identification (MFI) on 4 single polarization 32 GBaud signals: OOK, pulse-amplitude modulation (PAM4), binary phase-shift keying (BPSK), and quadrature phase-shift keying (QPSK) achieving an accuracy of >92% with an optical signal-to-noise ratio (OSNR) of 16 dB and up to >98% for an ideal OSNR and 20 km standard single mode fiber (SSMF) scenario [23]. We have also experimentally tested MFI on a set of four QAM formats (4QAM, 16QAM, 32QAM, 64QAM) of the same 32 GBaud symbol rate with an experimental NeuroPIC and achieved up to >97% accuracy for 7 dB OSNR after a 100 km long optical fiber transmission [24]. In both scenarios, there are four outputs for the digital readout, which is a secondary neural network (NN) after the reservoir itself, with a multiclass classifier to identify the input modulation format. Table 1 summarizes the state of the art along with the respective use cases and reported performances. Although a direct comparison is difficult, the variety of use cases demonstrates the broad scope of PRC capabilities. In addition, most of the reported works focus on simulation and numerical analysis, while only a limited subset of studies demonstrates practical implementations of physical photonic reservoir hardware.

*Table 1: The comparison of the literature*

| Paper | Design/Architecture | Task | Performance/Accuracy | Method |
|---|---|---|---|---|
| [1,10] | Coupled SOAs | Speech recognition | $> 97\%$ | Simulation |
| [5] | SiN / 4-Port architecture with passive splitter nodes | 64QAM detection | $< 1 \times 10^{-3}$ BER (100 km) | Simulation |
| [6] | Recurrent optical spectrum slicing neural networks (ROSS-NNs) | Mitigation of transmission impairments (IM/DD 112 GBaud PAM-4) | Signal equalization $> 100\ GBaud$ Signal transmission $> 60\ km$ | Simulation |
| [11] | Coupled SOAs | Classification of noisy time series | 96% (noise-free) 100% (Noisy) | Simulation |
| [12] | SOI / 4x4 swirl architecture with non-linear MRR nodes | Non-linear boolean operation (XOR) | $< 1 \times 10^{-3}$ BER (20 Gbps) | Simulation |
| [13] | InGaAsP-InP / 5x5 rainfall architecture with non-linear MRR nodes | Classification 8-bit non-return-to-zero (NRZ) patterns | 99.5% (160 Gbps) | Simulation |
| [14] | SOI / 4x4 swirl architecture with passive splitter nodes | Boolean operation (temporal XOR) | $< 1 \times 10^{-3}$ BER (32 Gbps) with reduced power consumption | Simulation |
| [19] | Directional coupler-based reservoir and particle swarm optimization readout | Optical channel distortion equalization on OOK signal | $9.15 \times 10^{-5}$ BER (20 Gbps, 25 km) | Simulation |
| [20] | Photonic crystal cavities | Up to 6-bit header recognition / boolean operation (XOR and AND) | $< 1 \times 10^{-3}$ ER (100 Gbps header recognition, 25-67 Gbps XOR) | Simulation |
| [21] | SOI / 4x4 swirl architecture with passive splitter nodes | Up to 5-bit header recognition / boolean operation (XOR) | 12.5 Gbps | Experimental |
| [22] | SiGe / 4-port architecture with 32 nodes | Mitigation of fiber non-linear distortion in 32 Gbps OOK | $< 0.2 \times 10^{-3}$ FEC $< 1 \times 10^{-3}$ BER | Experimental |

| | [24] and this work | SOI / 4x4 swirl architecture with passive splitter nodes | Modulation format identification | Up to 100% | Experimental |

Here, we investigate an experimental PIC implementation of the 4x4 node 4-port architecture experimental on our SOI photonics platform for MFI in the telecom C-band (see photo in Figure 1b. We compare the experimental NeuroPIC with a simulation in a system-level transceiver testbed. The experimental NeuroPIC outperforms its simulated counterpart in all the tested scenarios. This observation can be explained by a more complex output observed in the experimental NeuroPIC in Principal Component Analysis (PCA).

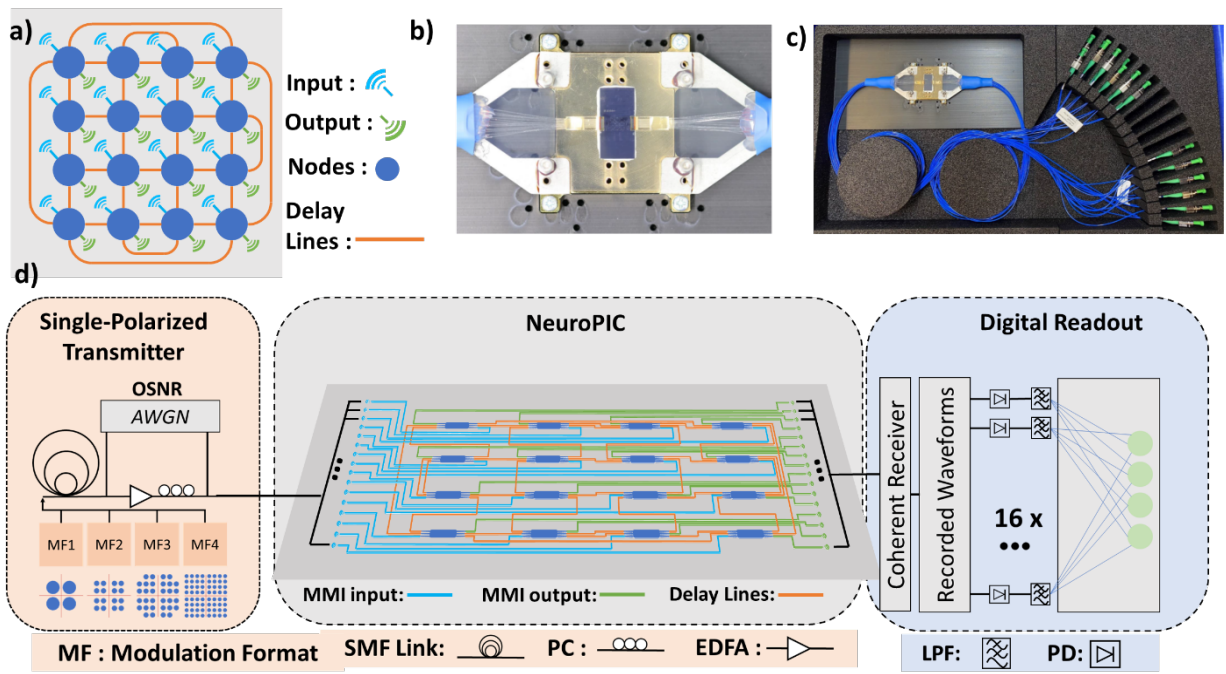

*Figure 1: a) Schematics of the 16 node 4-port architecture reservoir. b) Photo of the experimental NeuroPIC, c) Photo of the packaged NeuroPIC, d) PRC system test bed including incoming single-polarized transmitter, reservoir, and the final digital readout including the PD and the NN. (Abbreviations: OSNR: optical signal-to-noise ratio, AWGN: additive white Gaussian noise, SMF: single mode fiber, PC: polarization controller, BPF: band pass filter, PD: photodiode.)*

## 2. Methods

The NeuroPICs were designed using the Ansys Lumerical solvers to optimize the geometry of the components for minimal transmission losses. They were fabricated on our experimental SOI photonics platform and measured in testbed, including transmitters, the reservoir, and a digital readout (Figure 1d).

### 2.1. Design

The Ansys Lumerical Finite Difference Eigenmode (FDE) solver was employed to design the waveguide geometry to achieve modal confinement and propagation efficiency of the fundamental mode. We chose isolated shallow-etched rib waveguides for our NeuroPIC to minimize the waveguide losses from process-induced sidewall roughness. While it is generally important to minimize the loss of signal information and maintain a decent signal-to-noise ratio (SNR), PRCs specifically benefit from low losses because the resultant prolonged signal mixing within the reservoir provides richer dynamics of signal interference. This leads to a higher performance in RC [10].

We first describe the design of the individual components of our NeuroPIC. The rib waveguide core design measures 70 nm in height and 700 nm in width, while the slab spans 150 nm in height and 2.3 µm in width. A schematic cross-section of the waveguides is shown in Figure 2a-d and top SEM views are in Figure S1a. The reservoir was designed with 3x3 MMIs (Figure 2e), which serve as nodes, and 3.68 mm long waveguide delay lines that establish connections between these nodes. The delay lines are simply waveguides of specific lengths (Figure S1b) and play a crucial role by transforming the circulating signal into delayed versions of themselves at the next node, thereby providing short-term memory. They were designed to create a 47 ps signal delay between the nodes, which is comparable to the 31.25 ps symbol duration at 32 GBaud. This delay of 1.5 times of the symbol rate was sufficient to achieve a low misclassification rate [23]. The depth of this memory is constrained by all optical power losses. The MMIs distribute the light throughout the NeuroPIC. It was designed using the Ansys Lumerical Eigenmode Expansion (EME) solver with 3 inputs and 3 outputs that equally split the signal 3 ways, with 33% or -4.77 dB in each MMI output. The waveguide crossings (wgCrs) were designed in symmetric segments across all 4 ports with Ansys Lumerical 3D FDTD (Figure 2f). Their shape was optimized for maximum light transmission to opposite ports with minimum crosstalk into the adjacent orthogonal ports. All waveguides, MMIs, and wgCrs share the same rib waveguide structure to avoid tapers. A minimum bend radius of 40 µm was chosen to minimize radiation and mode mismatch losses, (Figure S1c). The light is coupled in and out of the NeuroPIC chip using uniformly pitched grating couplers (GCs). The GCs are optimized for coupling TE-polarized light with a wavelength of l = 1550 nm. Their critical dimension is 315 nm, resulting from 630 nm period with 50% duty cycle, (Figure 2g).

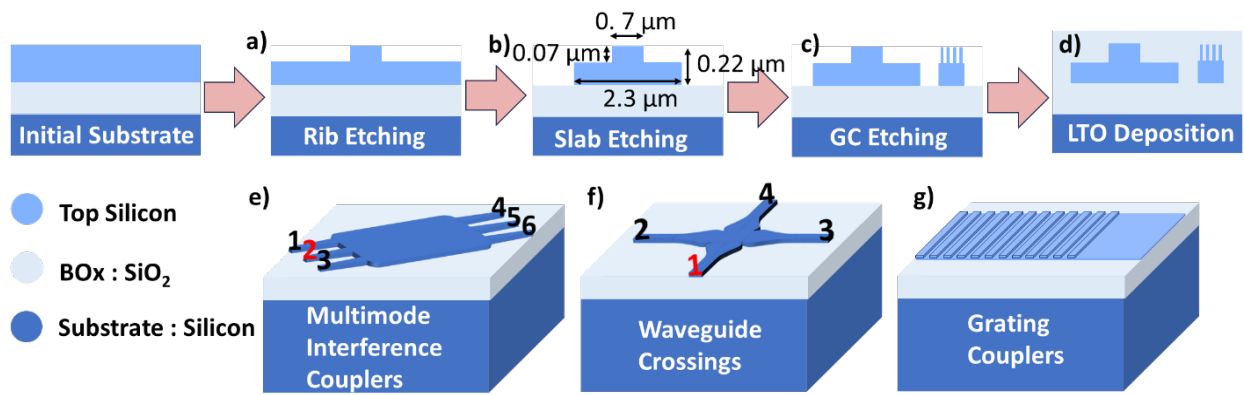

Figure 2: Schematics of the device fabrication, with cross-sectional view at the top and 3D views of the components in the bottom after slab etching. a) First i-line lithography with stepper and following RIE, b) second i-line lithography with stepper and following RIE, c) e-beam lithography and following RIE, d) LTO cladding deposition, e) multimode interference coupler (MMI) f) waveguide crossings (wgCr), g) grating couplers (GC).

The experimental basis for this work is the physical reservoir of our experimental NeuroPIC in a 4-port architecture, The light is coupled in and out of the reservoir through two ports of each MMI with input and an output GCs, respectively. The remaining MMI ports are linked to other MMI nodes inside the reservoir via the delay lines, as illustrated in Figure 1d. Overall, the reservoir is composed of 16 MMI nodes, resulting in 16 input and 16 output ports that are connected to the input/output GCs. This design ensures that all MMIs receive an equal input signal power. The GCs are strategically positioned along the sides of the reservoir to enable easy access by two 16x fiber arrays. This complexity and the planar 2D layout constraints introduced the need for 75 low-loss wgCrs. The investigations also include a comparison to a digital version of the physical reservoir, which we call the simulated NeuroPIC. Both versions of the NeuroPIC, fabricated and simulated, are connected to a digital readout and an NN classifier (Figure 1d).

### 2.2. Fabrication

We utilized our 150 mm SOI technology platform for the NeuroPIC fabrication, which adheres to CMOS fabrication constraints. The process starts with the WG core patterning step on 150 mm SOI wafers with 3 µm buried oxide (BOX) and 220 nm top silicon. We employed a Canon i5 i-line stepper to pattern positive tone AZ MiR701 resist. The waveguides were etched with a fluorine-based reactive ion etching (RIE) process in an Oxford Instruments Inductively Coupled Plasma (ICP)-RIE tool. The slab etching step followed a similar patterning and etching protocol. Figure 3a shows a cross-section scanning electron microscope (SEM) image of the fabricated waveguide.

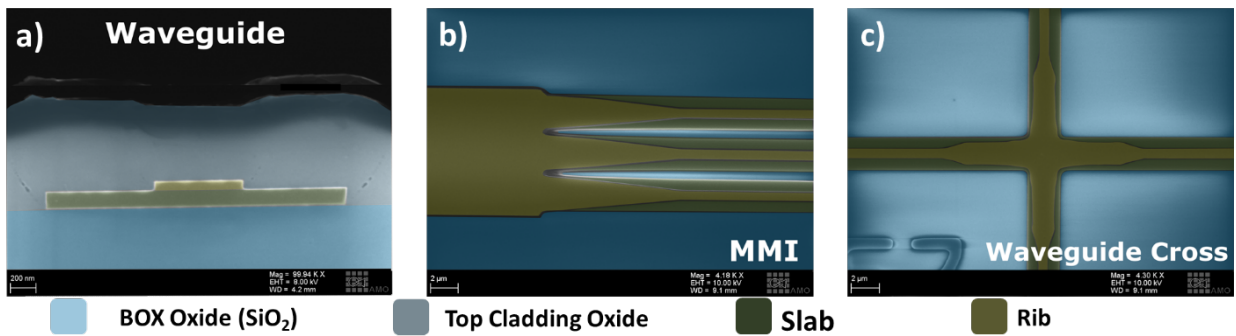

Figure 3b shows a top-view SEM image of the MMI output ports, and

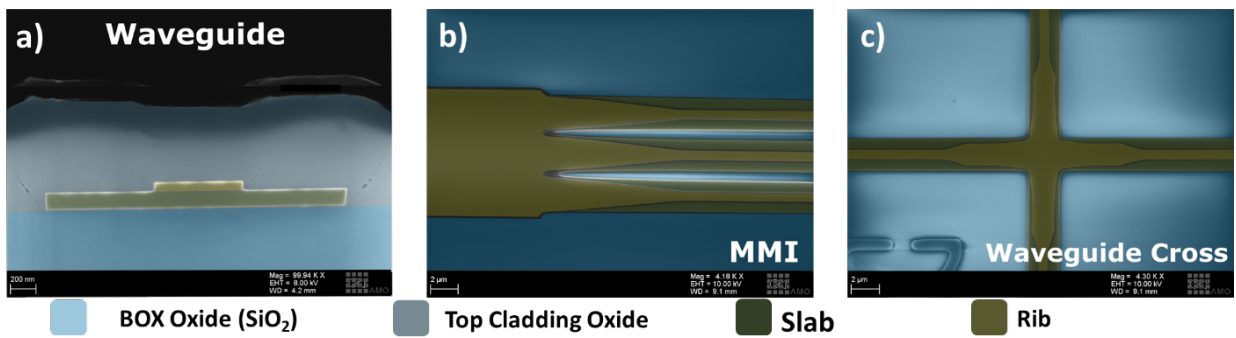

Figure *3*c shows a top view SEM image of the wgCr. The smaller critical dimensions of the GCs are lower than the resolution limit of the i-line stepper. They were, hence, patterned with electron beam lithography (EBL) using the same RIE process for etching. After resist removal, we employed standard RCA cleaning to remove inorganic and organic residues. Finally, we deposited a low-temperature oxide (LTO) film via LPCVD as cladding. The NeuroPIC is packaged with custom-made fiber arrays (see Figure 1c and supporting information Figure S2).

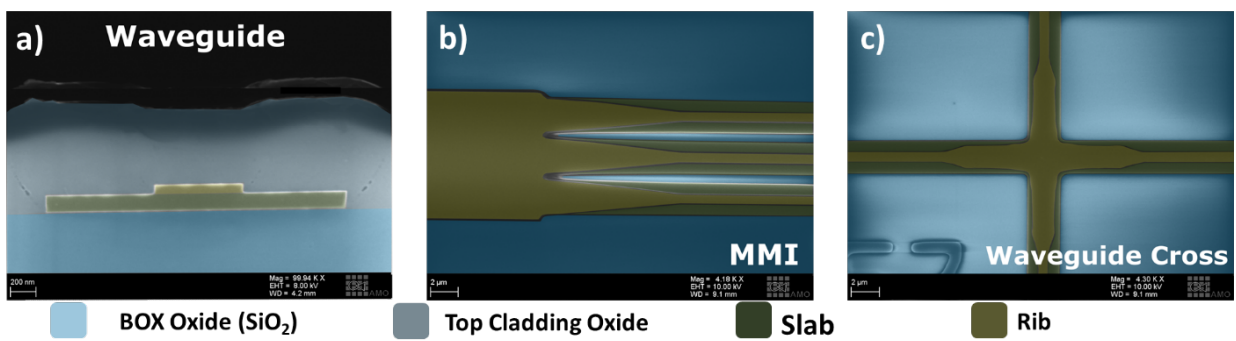

*Figure 3: (Colored) SEM images of a) the cross-section of the waveguide, b) the top view of the MMI ports and c) the top view of the waveguide cross.*

### 2.3. Measurement Setup

**Component-level characterization** was performed by placing the wafer between a single-mode input fiber and a multimode output fiber. We used a laser tunable in the wavelength range of λ = 1520 nm to 1620 nm. The input fiber was connected to the laser source, while the output fiber was connected to a photodetector. The open ends of these fibers were aligned with the input and output GCs of the components under test with an experimentally optimized angle of 13°. The output spectrum was recorded with a spectral resolution of 50 pm. Subsequently, the acquired spectra were analyzed using MATLAB to extract and quantify the waveguide propagation, MMI splitting and insertion, wgCr crosstalk and insertion, and GCs losses of the components.

**The system-level characterization** was conducted in a testbed that involves the generation and transmission of 32 GBaud single polarization signals, the reservoir NeuroPIC, the readout with a coherent receiver, photodiodes, and an NN classifier. The transmitter delivers optical signals with differently generated modulation formats applied on pseudorandom bit sequences through a SSMF fiber link of a specific distance to the photonic reservoir, where the complex interference pattern is generated at the outputs. The NeuroPIC system receives 32 GBaud single polarization signals that are generated by modulating a continuous wave (CW) signal at a wavelength of 1550 nm. This CW signal is produced using an external cavity laser. The modulation process is carried out by an in-phase quadrature modulator (IQ MOD), which is driven by a digital-to-analog converter (DAC). To amplify the signal, erbium-doped fiber amplifiers (EDFAs) are employed, and subsequently, the amplified spontaneous emission (ASE) noise is filtered out using an optical bandpass filter (OBPF). The 32 Gbaud data stream composed of different modulation formats was then split equally and distributed into the 16-channel fiber array, and then coupled into the input GCs on the NeuroPIC chip to fed into all reservoir nodes, with each GC receiving +4.4 dBm of power. The reservoir performs a transformation of the input, and the resulting output signal is optimized by adjusting the polarization controller (PC). These signals are received by the coherent receiver where the electro-optic conversion takes place. A subsequent digital readout is used to emulate 16 low-speed photodiodes (PD), one for each output, via a finite impulse response (FIR) low pass filter (LPF) of 100 MHz combined with down sampling for each output. This is where the nonlinearity is introduced and on which the training and testing are done. The outputs of the photodiodes are connected to a simple feed forward (FF) digital NN with a single fully connected hidden layer that is used to map the 16 PD outputs to 4 classifier outputs corresponding to the 4 modulation formats 4QAM, 16QAM, 32QAM, 64QAM to perform our final identification task. The single hidden layer is the only part of the NN that is trained, keeping the energy costs to a minimum. The schematic of the entire system is shown in Figure 1d. The output of the multi-classifier NN is realized using the "SoftMax" mathematical activation function, which exponentiates the input values and divides them by the sum of all exponentiated values. The result provides the probability of the input belonging to a specific class. This simple NN is sufficient to train the network and identify the modulation, thanks to the existence of the complex reservoir. The details of this approach are explained in our previous work [24].

## 3. Results

We first tested the individual optical components of the NeuroPIC on test wafers and used that data to design and fabricate the chips.

### 3.1. Optical Components Characterization

One of the key challenges in the NeuroPIC are optical losses from waveguide propagation, radiation, mode mismatches at bends, coupling during input and output through GCs, splitting at the MMIs, and wgCrs. These losses are caused to a large extend by scattering due to the sidewall roughness induced during the lithography and/or RIE etching processes.

The waveguide and GC losses were characterized by cutback measurements with 10 waveguides of incremental lengths ranging from 1.4 cm to 8 cm using a tunable C-band laser in the wavelength range between 1530 nm and 1570 nm. The measured spectra show maxima at around 1550 nm, in agreement with the GC design target (Figure 4a). The losses of each waveguide were plotted against the waveguide lengths. A propagation loss of 5.95 dB/cm was extracted from the slope of the resulting graph (Figure 4b). This value is below typical values of 1 dB/cm for such waveguides, but it is sufficient for the NeuroPIC, as chapter 3.2 will show. The y-intercept in the cutback plot in Figure 4b corresponds to a loss of 8.1 dB. We assume this value to be dominated by the coupling loss of both GCs, i.e. 4.05 dB/GC.

The 3x3 MMIs have been designed to evenly distribute incoming light from any of the three inputs to the three output ports. The transmission through the MMI test structures includes the GC coupling loss at both ends, and the waveguide propagation loss before and after the MMIs. The MMI transmission was, hence, calculated by subtracting the measured waveguide and GC losses. Figure 4c shows the comparison of experimental vs. simulated MMI transmission spectra of all 3 output ports for a single input to the MMI (port-2). The results are similar when using input port-1 and port-3, as shown in Figure S3 of the supporting information. The fluctuations in the measured spectra in Figure 4c is attributed to the grating couplers, as they are already present in the reference measurements and can only be partially compensated by calibration. Nevertheless, the MMIs work well in the desired wavelength range.

The wgCr transmission was calculated in a similar manner to that of the MMIs. Figure 4d further compares the wgCr transmission measurements with its simulation, also confirming that the wgCrs work well in the 1530 to 1570 nm band, which is consistent with the design targets and simulations. In addition, crosstalk measurements indicating the signal loss through the adjacent ports are shown in Figure S4.

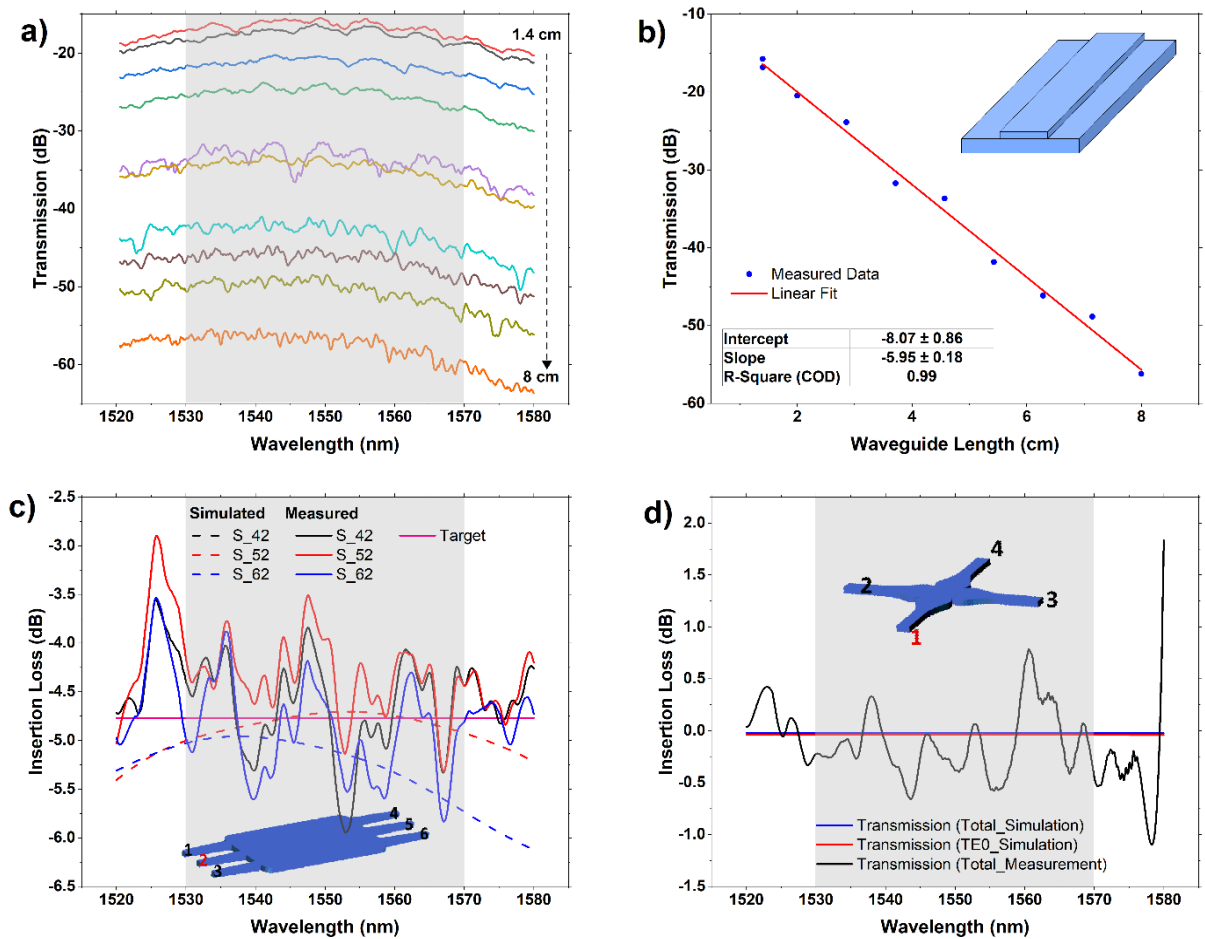

*Figure 4: Optical measurement results of the photonic reservoir components. a) The plot shows the transmission spectra for each waveguide length. b) Cutback plot: Insertion loss at 1550 nm versus waveguide length for ten different waveguide lengths. The slope of the curve represents the propagation loss, and the Y-intercept gives the coupling loss for the two GCs used for in/out-coupling. c) Insertion Loss spectra of MMI measurements (solid) vs. simulation (dashed) when the light was injected into input port-2 (inset: schematic of the 3x3 MMI with labeled*

*input and output ports). d) Insertion Loss spectra of wgCr measurements (black) vs. simulation (red and blue) when the light is injected to input port-1 (inset: schematic of a wgCr with labeled input and output ports).*

### 3.2. System Level Characterization

We conducted a series of experiments with the experimental NeuroPIC using 3 different symbol per sequence (SPS) lengths of 5120, 20480, and 40960 for a 20 km fiber link with a 32 dB OSNR for each modulation format. The temperature of the NeuroPIC was kept at 25°C during the measurements. The resulting data were then compared to data obtained with a simulated NeuroPIC, realized using Pytorch [25]. For the latter, the S-parameters of the components like MMI and wgCr were imported from the optical characterization of the test structures as described above (see [23,25] for details on the simulation).

The SPS length is a parameter that determines the length of the sequences with which the readout is trained to perform MFI. The value represents the length of a sequence of symbols with one modulation format, after which it has a probability of switching to other modulation schemes. High MFI accuracy is achieved for longer SPS lengths, whereas shorter SPS are more challenging. The learning process of the NN involves two essential steps: training and validation, which together form an epoch and are repeated over a certain number of epochs until MFI performance saturates at 100% or 100 epochs have passed. During an epoch of training and validation, the weights are updated as part of the learning process. Finally, the MFI testing begins using a distinct dataset, employing the best-performing model, typically the best-performing after 100 epochs.

The results of our experimental NeuroPIC are plotted in Figure 5a as a magenta star, as it performs at 100% accuracy for all SPS lengths. The simulated NeuroPIC, in contrast, shows the highest accuracy of 100% only for the longest 40960 SPS signals, and drops to 97% and 75% accuracy for 20480 and 5120 SPS, respectively. These numbers were extracted from Figure 5a at 6 dB/cm WG propagation loss to compare them to the experiment. The simulations further allowed exploring the parameter space, namely at different WG losses. The simulated MFI performance drops with higher WG losses, although the simulated NeuroPIC still achieves 100% accuracy for the longest 40960 SPS sequence at a high waveguide loss of 10 dB/cm. As a sanity check, we bypassed the NeuroPIC and fed the readout directly with the input data from the transceiver, resulting in a random guessing performance of 25%, even after optimizing the output NN. The experimental NeuroPIC is, therefore, necessary for MFI. It further performs significantly better than the simulated NeuroPIC and may thus be able to handle even more complex tasks. We believe the discrepancy between the simulated and experimental NeuroPICs is due to the inability of the simulation to recreate the variations from fabrication imperfections such as, phase, component-to-component deviations, etc. present in the experimental NeuroPIC.

We further investigated the discrepancy between the experimental and the simulated NeuroPIC through the MFI training evolution as a more sensitive performance measure than the saturated or final MFI performance after 100 training epochs (as reported in Figure 5a). The evolution of MFI was simulated and plotted against epochs in Figure 5b (dashed lines) using the experimental waveguide loss of 6 dB/cm and a learning rate of 0.01 (the same as used in Figure 5a) with varying SPS lengths. Shorter SPS lengths lead to slower learning progress and require a greater number of epochs to arrive at the final saturated MFI performance. However, the experimental NeuroPIC (solid lines) shows 100% accuracy from the start, which means that a single epoch is sufficient for training MFI. We further simulated the impact of waveguide losses on MFI with a challenging sequence length of 5120 SPS and a learning rate of 0.01 with different waveguide losses (1 dB/cm, 6 dB/cm, and 10 dB/cm; Figure 5c). We show that lower waveguide losses achieve higher MFI and faster training. The best case of 1 dB/cm achieves 85% and the worst case of 10 dB/cm achieves 73% accuracy. This observation cannot yet explain the discrepancy to 100% accuracy of the experimental NeuroPIC, but the results obtained with sub-optimal conditions in both experimental and simulated NeuroPICs (100% with a 6 dB/cm waveguide loss in the experimental PIC and 73% with a high 10 dB/cm waveguide loss in the simulation) demonstrate the reservoir's robustness against propagation losses.

We considered additional differences between the experimental and the simulated NeuroPICs and tested different extremes of random phase errors from fully random (0, 2π) to constant phase (0,0) on each delay line to mimic phase errors from fabrication imperfections. It should be noted that the phase error implemented is fully random in the previous plots. In Figure 5d, we show the MFI for different phase randomness levels at the delay lines, ranging from 0 to 2π. There is almost no noticeable difference or correlation with the randomness of the phase. The target length of each delay line in the NeuroPIC is $3.68 \times 10^{-3}$ m, corresponding to a time delay of 47 ps, which translates to a frequency of ~21 GHz. However, due to routing constraints, the mean length of the delay lines in the final layout is $3.7 \times 10^{-3}$ m with a standard deviation of $7.7 \times 10^{-5}$ m. We hence also investigated the effect of the delay line length randomness by implementing the standard deviation in the simulated NeuroPICs, where higher variations correspond to a larger anticipated degree of randomness. Figure 5e presents numerical results demonstrating MFI performance versus the standard deviation of delay line length for three different SPS in a 20 km fiber link with a 32 dB OSNR, similar to the example in Figure 5a. For longer SPS lengths, the simulation achieves 100% MFI until a delay length tolerance of $10^{-2}$ m, while shorter SPS lengths exhibit lower accuracy, as expected. Beyond a delay length tolerance of $10^{-2}$ m, the waveguide losses of the longer delay lines start to dominate. Consequently, the PIC loses its reservoir properties, and the accuracy drops to 25%, effectively rendering the system equivalent to random guessing. The MFI performance increases slightly and peaks for a delay line length standard deviation in the range of 1 to 10 mm. This is evident only for 5120 SPS and 20480 SPS as the less challenging scenario of 40960 SPS achieves 100% MFI accuracy for all standard deviations smaller than 10 cm. Although this shows that introducing randomness in the lengths of interconnecting delay lines moderately improves MFI accuracy, it cannot explain the observed differences between the experimental and the simulated NeuroPICs, as the fabricated one has a comparably low delay line length standard deviation of $7.7 \times 10^{-5}$ m. We repeated the evolution of the MFI with different delay line length standard deviations of $3 \times 10^{-5}$ m to $1 \times 10^{-1}$ m with two different waveguide losses of 6 dB/cm and 1 dB/cm. We used 5120 SPS length, where we can follow the performance change against the change in standard deviations as it is not saturated at 100% performance. The results are plotted in Figure 5f. Again, the performance increased up to standard deviations of ~$10^{-2}$ m, beyond which the networks (also) lose their accuracy reservoir property. This confirms that randomizing the delay lines improves the reservoir to some extent and that the performance is highly tolerant to the waveguide loss until at least the tested loss of 6 dB/cm. More data on the effect of different levels of delay line length standard deviation on the evolution of the MFI accuracy is provided in the supporting information Figure S5.

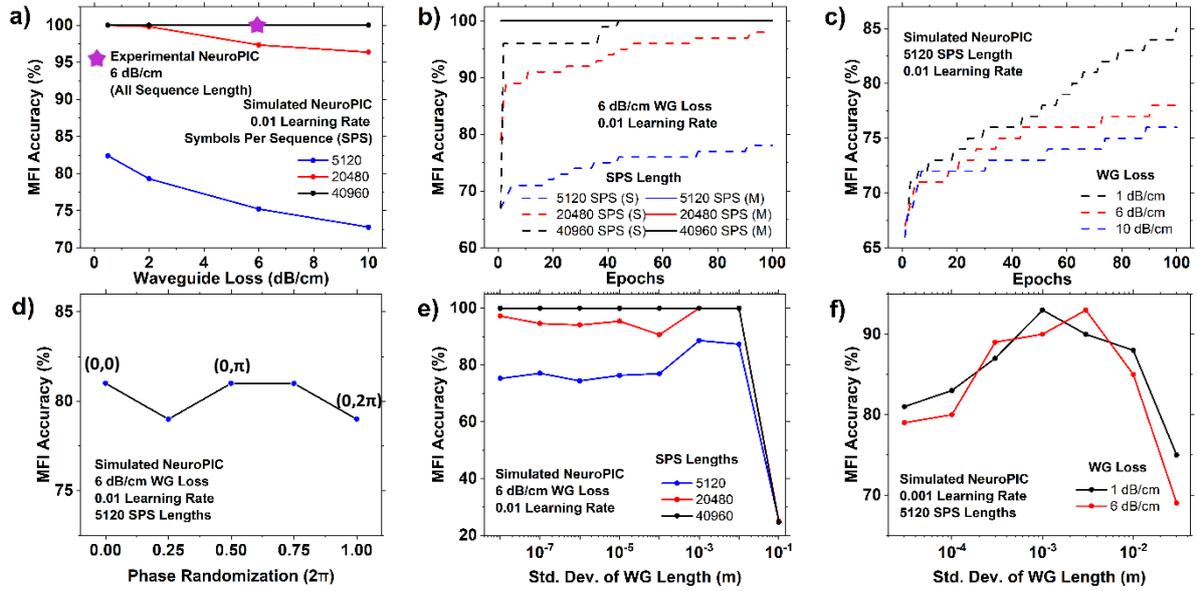

*Figure 5: a) MFI accuracy versus waveguide loss for different SPS lengths for simulated and experimental NeuroPIC, b) MFI accuracy versus number of epochs for different SPS lengths of experimental and simulated NeuroPIC, c) MFI accuracy versus number of epochs for different waveguide losses on simulated NeuroPIC, d) MFI accuracy versus different extremes of random phase accumulation in the delay lines (from (0,0) to (0,2π)), e) MFI accuracy versus standard deviation of waveguide lengths for different SPS length and f) MFI accuracy versus standard deviation of waveguide lengths for two different waveguide losses.*

Finally, we investigated the direct outputs of the 16-channel experimental and simulated NeuroPICs with a statistical technique known as Principal Component Analysis (PCA), a tool to reduce high dimensional data while preserving underlying trends and patterns. Figure 6a shows a comparison of the PCA results, plotting the eigenvalues against the number of principal components for both NeuroPICs. The output of the experimental and simulated NeuroPICs are reduced to ten and six principal components for the specific data we used for the MFI, respectively. This suggests that the output from the experimental NeuroPIC is more complex or "richer" in the language of reservoir computing. Figure 6b illustrates the cumulative variance versus the number of principal components used to explain it, leading to a similar conclusion. Although this dimensionality analysis reveals a direct behavioral difference between the experimental and simulated NeuroPICs, the origin of this difference is still under investigation.

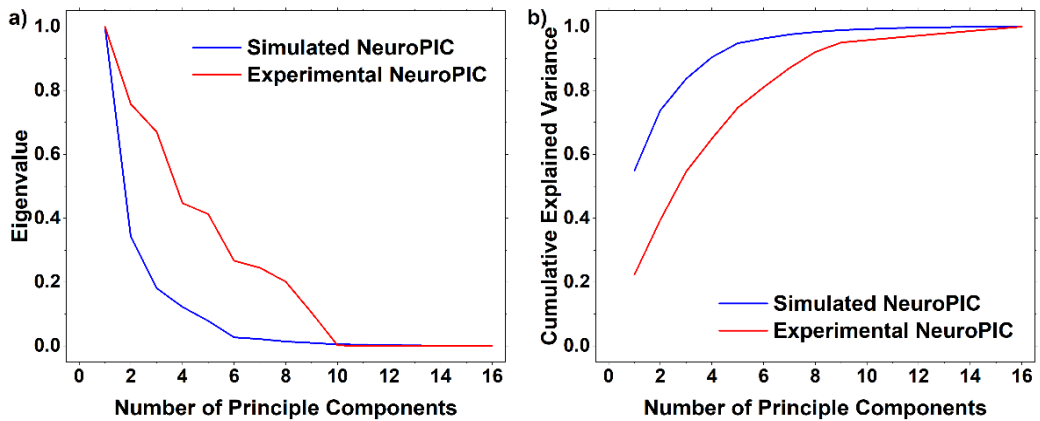

*Figure 6 : Principal component analysis for the 16 outputs of simulated NeuroPIC (blue) and experimental NeuroPIC (red). Number of principal components vs. a) Eigenvalue and b) cumulative explained variance.*

**4.  Conclusion**

We have designed and fabricated an experimental photonic reservoir computing circuit, NeuroPIC, and analyzed the use case of MFI in the telecom C-band. All the necessary optical components for photonic reservoir computing, such as waveguides, grating couplers, 3x3 MMIs, and waveguide crossings, were simulated, fabricated, and optically characterized individually. The NeuroPIC consisted of a 4x4 16-node 4-port reservoir architecture and was tested in a full system-level transceiver test bed. As part of this testing, 4 different modulation formats with randomly scrambled polarization were generated and transmitted over 20 km fiber links before being split into 16 input channels to each of the nodes of the NeuroPIC. The signals were allowed to mix freely inside the reservoir. The 16 outputs from each of the reservoir nodes were detected and passed to a digital readout section, after which a simple digital neural network was trained and tested. The modulation formats used for the training and testing were 4QAM, 16QAM, 32QAM, and 64QAM, with a symbol rate of 32 GBaud. The entire setup was also compared with a numerical implementation of NeuroPIC, subjected to scenarios of different symbol sequence lengths, different waveguide losses of up to 10 dB/cm, a slower learning rate, extra phase randomization and varying amounts of randomness in the delay line length. The MFI accuracy of the system with the experimental NeuroPIC was ~100% in all our tested cases. Our simulations showed that the accuracy is generally expected to improve for longer sequence lengths, lower waveguide propagation losses and more randomness in the delay line lengths, with a high tolerance against waveguide losses. The experimental NeuroPIC performed better than any numerically predicted model, which we attribute to more complex or "richer" output on the fabricated device. The fact that 100% accuracy was achieved in all three tested scenarios, even with relatively lossy waveguides, is promising for solving more complex tasks in future applications.

**Acknowledgements**

Funded by the Federal Ministry of Education and Research of Germany (BMBF) in the framework of CELTIC-NEXT AI-NET PROTECT [project numbers FKZ 16KIS1291 (AMO GmbH), FKZ 16KIS1281 (Fraunhofer-Institut für Nachrichtentechnik, Heinrich Hertz Institute), and FKZ 16KIS1301 (ID Photonics GmbH)], under the 6G-RIC project with grant 16KISK020K and in the the Cluster4Future NeuroSys (FKZ 03ZU1106BB). The authors would like to express their sincere gratitude to Piotr Cegielski (now with Infineon Technologies, Munich), Anna Lena Schall Giesecke (now with University of Duisburg-Essen and Fraunhofer IMS, Duisburg) and Peter Bienstman (Ghent University) for their invaluable insights and stimulating discussions on the subject matter of this research.

**Supporting Information**

The following data are available free of charge in the SI (DOC):

SI1 – Figure S1: (Colored) SEM images of the a) waveguides b) delay lines c) bends.

SI2 – Description of the NeuroPIC Packaging, including Figure S2: Schematic and photograph of packaged NeuroPIC.

SI3 – Description and Data of further multimode interference splitters (MMIs), including Figure S3: Transmission of MMI light injection to port-1 and port-3.

SI4 – Description and transmission spectra (Figure S4) of waveguide crossings.

SI5 – Information on the influence of delay line length standard deviations on the evolution of MFI accuracy.

# Supporting Information

# Hardware Realization of Neuromorphic Computing with a 4-Port Photonic Reservoir for Modulation Format Identification

Enes Şeker[1,2], Rijil Thomas[1], Guillermo von Hünefeld[3,4], Stephan Suckow[1,*], Mahdi Kaveh[3], Gregor Ronniger[3], Pooyan Safari[3], Isaac Sackey[3], David Stahl[5], Colja Schubert[3], Johannes Karl Fischer[3], Ronald Freund[3,4], and Max C. Lemme[1,2,*]

(1) AMO GmbH, Advanced Microelectronic Center Aachen (AMICA), Otto-Blumenthal-Straβe 25, 52074 Aachen, Germany

(2) Chair of Electronic Devices, RWTH Aachen University, Otto-Blumenthal-Straße 25, 52074 Aachen, Germany

(3) Fraunhofer Institut für Nachrichtentechnik, Heinrich Hertz Institute (HHI), Einsteinufer 37, 10587 Berlin, Germany

(4) Technical University of Berlin, Photonic Communication Systems, Straße des 17. Juni 135, 10623 Berlin, Germany

(5) ID Photonics GmbH, Anton-Bruckner-Straße 6, 85579 Neubiberg, Germany

*Email: suckow@amo.de; max.lemme@rwth-aachen.de

**SI1: Component Design**

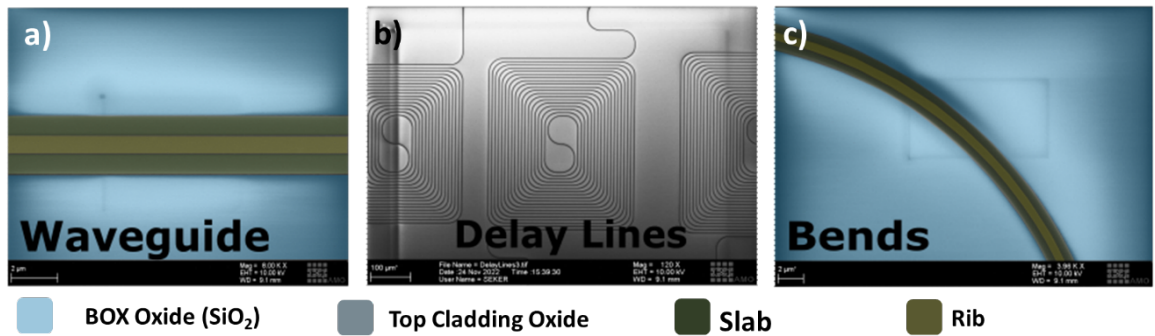

*Figure S1: (Colored) SEM images of the a) waveguides b) delay lines c) bends.*

**SI2: NeuroPIC Packaging**

To facilitate characterizing the experimental NeuroPIC, we outsource packaging of the experimental NeuroPIC. The NeuroPIC was securely affixed to a Peltier element, enabling precise temperature control during measurements. Featuring 16 reservoir inputs and 2 self-loop inputs for active alignment, a custom-made fiber array with an 18-channel fiber array ribbon, with polished angles meticulously set to maximize coupling efficiency to the grating couplers on chip. In the NeuroPIC design phase, additional alignment waveguides were strategically placed on each side of the chip, incorporating GCs at both ends. The two outer fiber channels were utilized for active alignment during the packaging process. The packaging ensures, efficient temperature control, and overall reliability of the NeuroPIC during measurements. Figure S4 shows the schematics and photo of the packaged NeuroPIC.

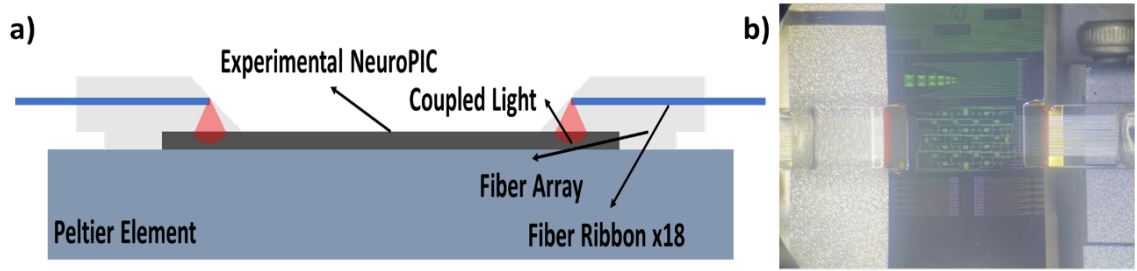

*Figure S 2:a) The schematics of the NeuroPIC packaging b) A photograph of packaged NeuroPIC.*

**SI3: MMI Performance**

An optimized 3x3 MMI ensures equal light distribution in all three directions. This implies that when light is injected into any of the ports, it is evenly divided among the outer ports. The paper already provides information about light injection into the second port, which resulted in an output spectrum centered at 1550 nm.

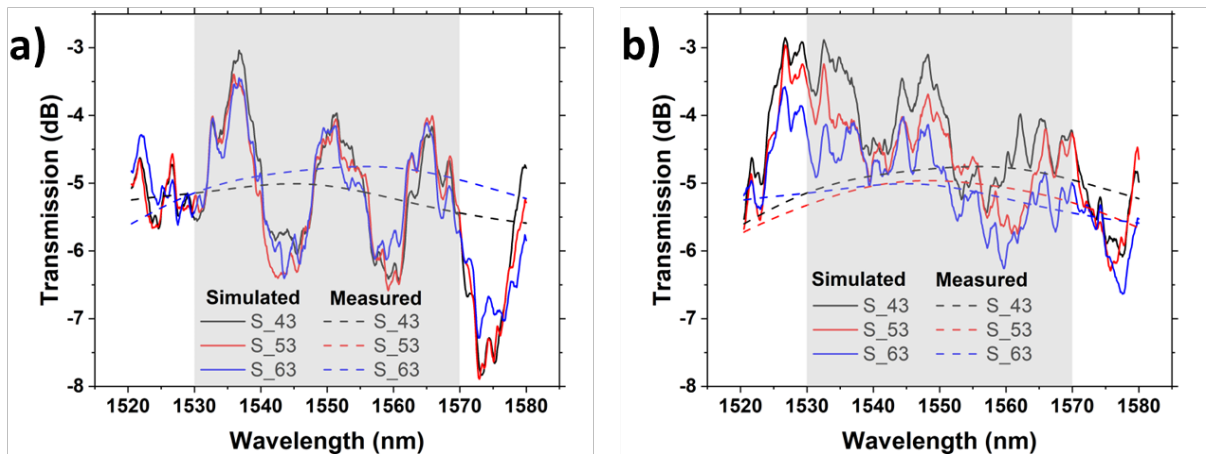

*Figure S3: Transmission of MMI light injection to a) port-1 and b) port-3. Dashed lines represent the simulation results while solid lines show the experimental results.*

**SI4: Waveguide Crossings**

Waveguide crossings (wgCr) possess four arms by design, which are optimized for minimal loss in the direct transmission between opposing arms. The most significant loss occurs due to light propagation to the neighboring arms, which is commonly referred to as crosstalk. Figure S4 displays the crosstalk measurements of the optimized wgCr's, demonstrating nearly negligible propagation to the adjacent arms.

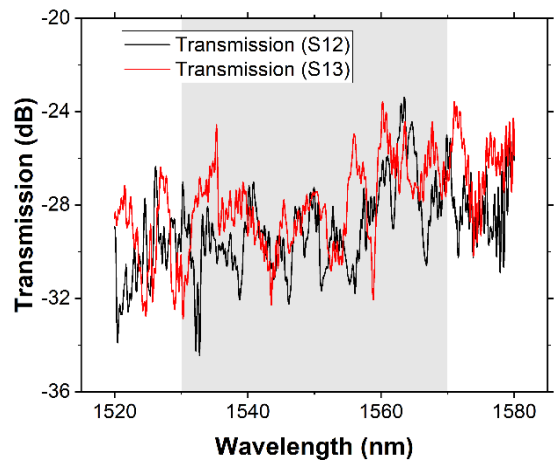

*Figure S4: Transmission spectra of wgCr`s on the neighbor arms, while injecting through port 1 and measuring from port 2 (black lines) and port 3 (red lines).*

**SI5: NeuroPIC Analysis**

Figure S 5 compares the effect of different levels of delay line length standard deviations ($3 \times 10^{-5}$ m to $1 \times 10^{-1}$ m) on the evolution of MFI accuracy.

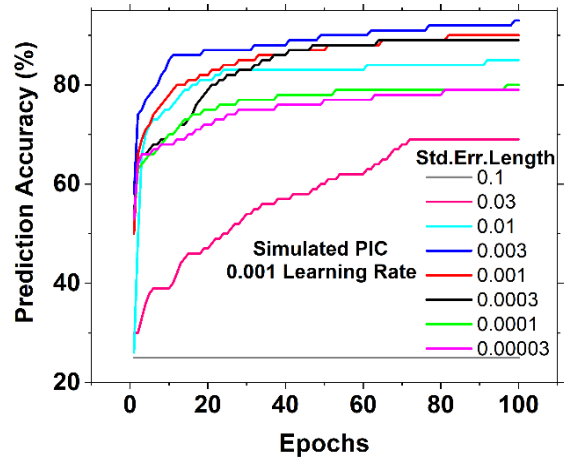

*Figure S 5: MFI accuracy versus number of epochs for different standard deviations of waveguide length.*

**Graphical Table of Content**

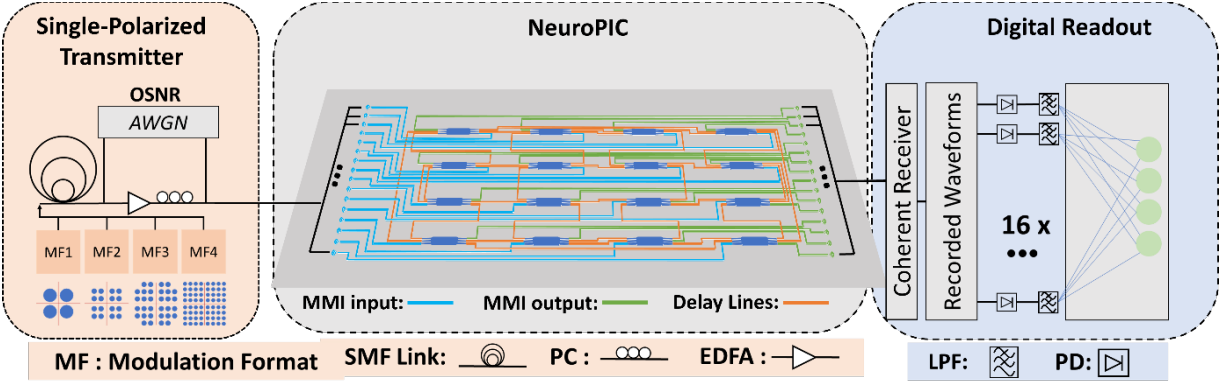

**Keywords:** photonic integrated circuits, neuromorphic computing, reservoir computing, neuromorphic PIC, telecom network monitoring, modulation format identification